\documentstyle[epsfig, twocolumn]{iso99}

\newcommand{\thepapername}{}

\begin{document}

\setlength{\parindent}{0pt}
\setlength{\parskip}{ 10pt plus 1pt minus 1pt}
\setlength{\hoffset}{-1.5truecm}
\setlength{\textwidth}{ 17.1truecm }
\setlength{\columnsep}{1truecm }
\setlength{\columnseprule}{0pt}
\setlength{\headheight}{12pt}
\setlength{\headsep}{20pt}
\pagestyle{veniceheadings}

\title{\bf CIRRUS SPECTRA OF LOW SURFACE BRIGHTNESS REGIONS
}

\author{{\bf M.~Juvela$^1$, K.~Mattila$^1$, D.~Lemke$^2$} \vspace{2mm} \\
$^1$Helsinki University Observatory, P.O.\,Box 14, SF-00014 University of
Helsinki, Finland \\
$^2$Max-Planck-Institut f\"ur Astronomie, K\"onigstuhl 17, D-69117 Heidelberg, Germany
}

\maketitle

\begin{abstract}

We have studied the galactic cirrus in low surface brightness regions using
ISOPHOT raster maps at 90\,$\mu$m, 150\,$\mu$m and 180\,$\mu$m. Observations
are used to determine dust emission spectra and dust temperatures. The data
extend to longer wavelengths than IRAS observation. Compared with DIRBE data
the resolution and the sensitivity are better and this makes it possible to
study faint cirrus emission in individual fields.

We will discuss the calibration of the observations and present results of a
comparison between ISOPHOT and DIRBE surface brightness values. The
correspondence was found to be better than $\sim$30\%. At 90\,$\mu$m the
ISOPHOT surface brightnesses tend to be slightly higher than the DIRBE
values while at longer wavelength the situation is reversed.

Surface brightness variations caused by cirrus fluctuations make it possible
to determine the spectrum of the dust emission. Cirrus spectra were obtained
for six fields with surface brightnesses in the range 1-2\,MJy\,sr$^{-1}$
after the Zodiacal light and the contribution of extragalactic sources have
been subtracted.

Assuming $\nu^2$ emissivity law the dust temperatures are in the range
18--20\,K. Temperature variations can be seen even within individual fields.
These values are higher than values from DIRBE but this can be explained by
differences in the calibration.
\vspace {5pt} \\


Key~words: ISO; infrared astronomy; cirrus.

\end{abstract}

\section{INTRODUCTION}

At wavelengths longer than $\sim100$\,$\mu$m the far-infrared emission of
the Galaxy is dominated by thermal emission from large dust particles that
are in thermal equilibrium with the interstellar radiation field. Dust
temperatures determined from DIRBE measurements are typically 16--19\,K
(e.g.\ \cite{mj_lagache98}; \cite{mj_dwek97}) in rough agreement
with model predictions (e.g.\ \cite{mj_draine85}).

Most previous FIR dust temperature determinations have been based on COBE
observations. These have, however, poor spatial resolution and in low surface
brightness regions the sensitivity of the detectors becomes a limiting factor
so that only an average dust spectrum over a large area can be determined.

We use ISOPHOT observations to study the spectrum of galactic cirrus in
regions selected for their low surface brightness. Because of the better
sensitivity and angular resolution of the ISOPHOT instrument it is possible
to determine the cirrus spectrum in these fields based on the cirrus
intensity variations. In this way constant or slowly varying components,
e.g.\ Zodiacal light, can be eliminated. 

The determination of the cirrus spectrum is important also for extragalactic
studies where the galactic foreground emission must be separated. In future
work the same fields will be used to estimate the FIR component of the
cosmic background radiation.

We start by presenting the observations used in this study. After that
calibration differences between ISOPHOT and DIRBE observations are
discussed. Calibration issues are obviously essential for the determination
of the spectra and the dust temperatures. Finally, the method used to derive
the cirrus spectra and the derived dust temperatures are presented. A
$\nu^2$ dust emissivity law will be assumed throughout this paper.

\section{OBSERVATIONS}

The cirrus spectra were determined in six fields that had been observed at
90\,$\mu$m, 150\,$\mu$m and 180\,$\mu$m with the ISOPHOT instrument
(\cite{mj_lemke}) aboard the ISO satellite (\cite{mj_kessler}). The observation
consist of PHT22 raster maps. Detailed description of the observations is
given in Table~\ref{mj_table:maps}. Data reduction was done using the
ISOPHOT Interactive Analysis software package (PIA) (\cite{mj_gabriel97})
version 8.0. In the following the analysis is based on measurements reduced
to the AAP level.

Flatfielding was done outside PIA using custom routines that also removed
slow drifts of the detector pixels relative to each other. Good flat
fielding is not critical for the determination of the spectra but it will
reduce the noise in the subsequent steps.

\begin{table*}
\caption[]{Description of the ISOPHOT maps used to determine the cirrus
spectra. The columns are: (1) name of the field, (2)-(3) coordinates of the
centre of each field, (4)-(5) galactic coordinates, (6) map area, (7)
ISOPHOT filter, (8) number of raster positions, (9) step between adjacent
raster positions and (10) integration time. The distance of adjacent scans
was the same as the raster step used along the scan line.}
\label{mj_table:maps}
  \begin{center}
    \leavevmode
    \footnotesize
\begin{tabular}{lrrrrrrrrr}
\hline \\[-5pt]
Field  &  \multicolumn{2}{c}{Map Centre} &  &   & 
                  Area & Filter  &  Rasters  & Step & $t_{\rm int}$  \\
      &    RA(2000.0) & DEC(2000.0)     & $l$ & $b$ & 
      (square degrees) &         &         &  (arcsec)  & (s)  \\
(1) & (2) & (3) & (4) & (5) & (6) & (7) & (8) & (9) & (10) \\      
\hline \\[-5pt]
VCN    & 15 15 21.7 &  +56 28 58  & 91.76 & 51.42 &  0.030
                            & C\_90  &  10$\times$4  &  90 & 46  \\
       &            & & & & & C\_135 &  10$\times$4  &  90 & 46  \\
       &            & & & & & C\_180 &  10$\times$4  &  90 & 46  \\	

VCS    & 15 15 53.1 & +56 19 30   & 91.27 & 51.40 &  0.023
                            & C\_90  &  21$\times$2  &  90 & 46  \\
       &            & & & & & C\_135 &  21$\times$2  &  90 & 46  \\
       &            & & & & & C\_180 &  21$\times$2  &  90 & 46  \\	

NGPN   & 13 43 53.0 & +40 11 35   & 86.82 & 73.61 & 0.27
                            & C\_90  &  32$\times$4  &  180 & 23 \\
       &            & & & & & C\_135 &  32$\times$4  &  180 & 27 \\
       &            & & & & & C\_180 &  32$\times$4  &  180 & 27 \\
       & 13 42 32.0 & +40 29 06   & 87.88 & 73.26 & 0.53
                            & C\_180 &  15$\times$15 &  180 & 32 \\	

NGPS   & 13 49 43.7 & +39 07 30   & 81.49 & 73.30 & 0.27
                            & C\_90  &  32$\times$4  &  180 & 23 \\
       &            & & & & & C\_135 &  32$\times$4  &  180 & 27 \\
       &            & & & & & C\_180 &  32$\times$4  &  180 & 27 \\	

EBL22  & 02 26 34.5 & -25 53 43   & 215.78 & -69.19 & 0.19
                            & C\_90  &  32$\times$3  &  180  & 23 \\
       &            & & & & & C\_135 &  32$\times$3  &  180  & 27 \\
       &            & & & & & C\_180 &  32$\times$3  &  180  & 27 \\	

EBL26  & 01 18 14.5 & 01 56 40    & 135.89 & -60.66 & 0.27
                            & C\_90  &  32$\times$4  &  180  & 23 \\
       &            & & & & & C\_135 &  32$\times$4  &  180  & 23 \\
       &            & & & & & C\_180 &  32$\times$4  &  180  & 23 \\	

%
\hline \\[-5pt]
\end{tabular}
\end{center}
\end{table*}

\section{COMPARISON WITH DIRBE SURFACE BRIGHTNESS VALUES} \label{mj_sect:cal}

The results presented in this paper are based on the calibration performed
with Fine Calibration Source (FCS) measurements. The absolute calibration of
ISOPHOT observations has been estimated to be better than 30\%
(\cite{mj_klaas98}). \cite*{mj_klaas99} recently reported for both C100 and
C200 accuracies better than 20\% relative to DIRBE.  These results apply to
fields with surface brightness above $\sim$5\,MJy\,sr$^{-1}$ while several
of our C200 maps are below 4\,MJy\,sr$^{-1}$. It is therefore useful to
compare our observations with DIRBE data especially as the calibration
accuracy is essential for the determination of the cirrus spectra.

The DIRBE and ISOPHOT calibration was studied in the case of the fields
NGPS, NGPN, EBL22, and EBL26 (see Table~\ref{mj_table:maps}). Three methods
were used to perform the comparison:

\begin{enumerate}
\item {\em Comparison of absolute surface brightness levels}. The ISOPHOT maps
were compared with DIRBE ZSMA ({\em Zodi-Subtracted Mission Average}) data
to which Zodiacal light was added according to the model given by 
\cite*{mj_leinert98}). Using the DIRBE weekly maps observed with the
same solar elongation as the ISO data did not change the results
significantly. Because of the lower noise we chose to use the ZSMA data.
Both DIRBE ZSMA and ISO were colour corrected assuming a $\nu^2 B_{\nu}$
spectrum with $T_{\rm dust}$=18\,K. The zodiacal light estimates were colour
corrected assuming black body spectrum with $T_{\rm dust}$=270\,K
(\cite{mj_abraham1}; \cite{mj_abraham2}). Corresponding curves fitted to the
DIRBE ZSMA 100\,$\mu$m, 140\,$\mu$m and 240\,$\mu$m data and the zodiacal
light estimates were used to derive the DIRBE surface brightness estimates
interpolated to the wavelength of the ISO observations. The average ISO flux
density was compared with the derived DIRBE values calculated as the average
over ISO map and weighted by the DIRBE beam. The result is the surface
brightness ratio $S$(DIRBE)/$S$(ISOPHOT).

\item {\em Direct comparison of the surface brightness variations} in ISOPHOT
and DIRBE ZSMA maps. All data were first colour corrected for a spectrum
$\nu^2\,B(\nu)$ with $T_{\rm dust}$ =18.0\,K. For each DIRBE pixel the data
were interpolated to the wavelength of the ISOPHOT observations and the
corresponding ISOPHOT surface brightness estimate was calculated as a
weighted average over the DIRBE beam. Linear fit was done to these points to
derive the slope between the surface brightness values 
$k=\Delta S$(DIRBE)/$\Delta S$(ISOPHOT).

\item {\em Comparison via IRAS}. Linear relations were established
between the IRAS 100\,$\mu$m data and the ISO observations and between the
IRAS data and the DIRBE ZSMA data interpolated to the wavelength of the ISO
observations using a fitted curve $\nu^2 B_{\nu}$($T$=18K). For each IRAS
ISSA map pixel inside the field the corresponding ISO surface brightness was
calculated as an average weighted with a Gaussian with FWHM$\sim$5 arcmin.
The relation between IRAS and DIRBE surface brightnesses was determined with
a similar method but over an area with radius $\sim$2 degrees where for each
DIRBE pixel the average IRAS value was calculated by weighting with the
DIRBE beam. The slopes of these two linear relations $\Delta S({\rm
DIRBE})$/$\Delta S({\rm IRAS(100\mu{\rm m})})$ $\Delta S({\rm IRAS(100\mu{\rm
m})})$/$\Delta S({\rm ISOPHOT})$ give the ratio between DIRBE and ISO scales,
$k=\Delta S$(DIRBE)/$\Delta S$(ISOPHOT).

\end{enumerate}

Method (1) suffers from the small size of the ISOPHOT maps which means that
for each DIRBE pixel a large portion of the flux comes from outside the area
mapped with ISOPHOT. 

Method (2) is even more affected by the limited map sizes and the large size
of the DIRBE beam since we must determine the relation between brightness
variations. However, NGPN and NGPS form together a 3.1 degrees long strip
where the surface brightness gradient is along the longer side of the map.
Here the fact that the ISO map is narrow compared with the DIRBE beam should
not lead to significant errors. On the other hand, in EBL26 the surface
brightness drops quickly outside the bright region in the northern part of
the map. The surface brightness variation is not resolved by DIRBE beam and
the values $\Delta({\rm DIRBE})$/$\Delta({\rm ISOPHOT})$ obtained would be
underestimated. In EBL22 and NGPN the method does not work because of the
lack of sufficient surface brightness variations.

Methods (2) and (3) are not affected by the presence of zodiacal light. The
last method uses IRAS ISSA maps as an intermediate step and is therefore
much less affected by the poor resolution of the DIRBE data. It allows also
the use of DIRBE data over a much larger area and thereby reduces the effect
of the large noise present in 140\,$\mu$m and 240\,$\mu$m DIRBE
observations. In these low surface brightness regions the error estimates of
140\,$\mu$m and 240\,$\mu$m DIRBE measurements exceed 50\%. On the other
hand, the method is based on the assumption that the ratio of the emission
at 100\,$\mu$m and at other wavelengths remains constant.

\begin{table*}
\caption[]{The ratios $k$ between the DIRBE and ISO flux density scales.
The columns are: (1) the name of the field, (2) wavelength of ISO
observations, (3) mean surface brightness of the ISO map, (4) $k$ from the
comparison of absolute surface brightnesses, (5) $k$ obtained from direct
comparison of surface brightness variations and (6) $k$ obtained via
100\,$\mu$m ISSA maps.  Formal errors from the least squares fits are shown in
parentheses. For the first method (i.e.\ column 4) the numbers given in
parentheses give the dispersion in the DIRBE surface brightness values
interpolated to the wavelength of the ISO observations }
\label{mj_table:dirbe}
  \begin{center}
    \leavevmode
    \footnotesize
    \begin{tabular}{lccccc}
    \hline \\[-5pt]
Map     &  $\lambda$
        & $<S>$  
	&  $S({\rm DIRBE})/S({\rm ISOPHOT)}$  
	&  $\Delta S$(DIRBE)/$\Delta S$(ISO) 
	&  via IRAS
	\\
        & ($\mu$m) &  (MJy/sr) &      &    &           \\
(1) & (2) & (3) & (4) & (5) & (6) \\
\hline \\[-5pt]
EBL22 
& 90   & 5.6     &  0.81(0.01)  &        &      \\
& 150  & 3.4     &  1.28(0.03)  &        &      \\
& 180  & 2.7     &  1.42(0.05)  &        &      \\

EBL26
& 90  & 14.3     &  0.84(0.02)  &     &  0.88(0.10) \\
& 150 & 9.1      &  1.04(0.05)  &     &  1.09(0.12) \\
& 180 & 7.4      &  1.09(0.03)  &     &  1.27(0.14) \\

NGPS
& 90   & 5.7     &  0.69(0.06)  &  0.69(0.10)   &  0.70(0.05)  \\
& 150  & 3.7     &  0.94(0.15)  &  1.68(0.15)   &  1.43(0.11)  \\
& 180  & 3.1     &  1.04(0.02)  &  1.59(0.15)   &  1.40(0.10)  \\ 

NGPN
& 90   & 5.0     &  0.67(0.02)  &  &  \\
& 150  & 3.6     &  0.97(0.04)  &  &  \\
& 180  & 2.7     &  1.19(0.11)  &  &  \\
& 180$^1$ & 2.6  &  1.21(0.07)  &  &  \\

NGPN \& NGPS
& 90   & 5.2     &  0.67(0.02)  &  0.63(0.21)  &  0.75(0.06)  \\
& 150  & 4.1     &  0.92(0.04)  &  0.91(0.34)  &  1.56(0.13)  \\
& 180  & 3.3     &  1.08(0.12)  &  0.84(0.33)  &  1.53(0.11)  \\

\hline \\[-5pt]
\end{tabular}

$^1$ the larger 180\,$\mu$m map \\
\end{center}
\end{table*}

The results are given in Table~\ref{mj_table:dirbe}. It can be seen that
DIRBE and ISOPHOT surface brightness values agree typically to within 30\%.
However, at 90\,$\mu$m the ISOPHOT surface brightness tends to be above the
value predicted from DIRBE observations while at longer wavelengths the
situation is reversed. The extrapolation of the DIRBE values to 90\,$\mu$m
using the modified black body fitted to three longer wavelength may not be
entirely justified and may lead to underestimation of the surface
brightness. Also, in these faintest regions the accuracy of the DIRBE
140\,$\mu$m and 240\,$\mu$m is poor and a reliable comparison with ISOPHOT
C200 values is difficult. The modified black body curves fitted to DIRBE
data are determined mostly by the 100\,\,$\mu$m DIRBE value which has
significantly smaller error estimates than the two longer wavelengths, and
the assumed dust temperature, $T$=18.0\,K. Increasing the temperature by one
degree would increase $k$ at 90\,\,$\mu$m by some 5\% and decrease it at
longer wavelengths by some 15\%.

In spite of uncertainties the results suggest that the dust temperatures
derived from ISOPHOT observations are, because of calibration differences,
higher than those based on DIRBE data. Using PIA versions prior to 8.0 the
differences between ISOPHOT and DIRBE surface brightnesses were typically
larger by some 20\%.

\begin{figure}[ht]
\begin{center}
\leavevmode
\centerline{\epsfig{file=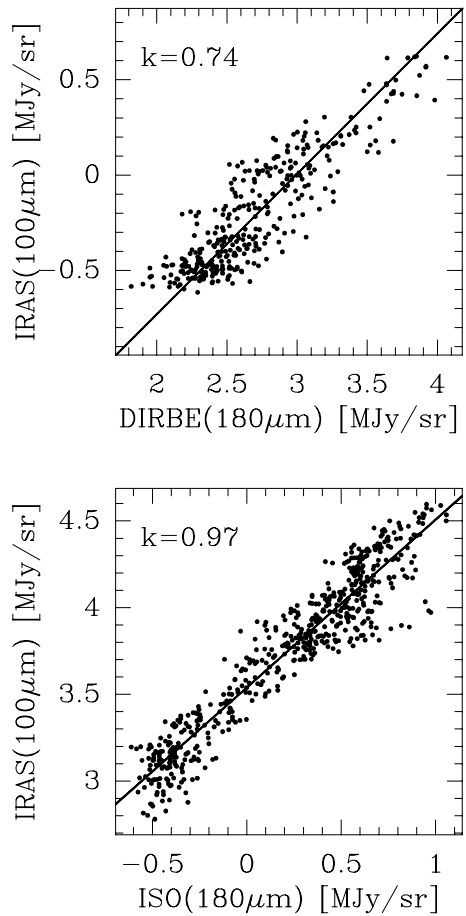, width=6.0cm}}
\end{center}
\caption{\em Determination of the relative scales of the
ISOPHOT and DIRBE maps using IRAS 100\,$\mu$m ISSA maps as an intermediate
step. The ISOPHOT map is the 180\,$\mu$m map of the field NGPS. The upper
frame shows linear fit between IRAS 100\,$\mu$m surface brightness values
and the DIRBE values interpolated to 180\,$\mu$m. The lower frame shows the
corresponding fit between IRAS 100\,$\mu$m and ISOPHOT maps. The slopes of
the fitted lines (shown in figures) can be combined to get the value shown
in column 5 of Table~\ref{mj_table:dirbe}.}
\label{mj_fig:via}
\end{figure}


\section{CIRRUS SPECTRA}

As the first step all data were colour corrected in order to derive
monochromatic surface brightness values at the reference wavelengths of the
filters. Correction was done for spectrum $\nu^2 B_{\nu}(T=18K)$ which
approximates the expected cirrus spectrum. A large fraction of the observed
surface brightness values is due to Zodiacal light. Since we will study only
the relative surface brightness variations the Zodiacal light is eliminated
from the analysis and it does not affect the required colour correction.

In the calculations the 180\,$\mu$m maps are used as the reference. For each
180\,$\mu$m pixel the corresponding surface brightness value at another
wavelength is calculated as a weighted average. The spatial weighting is
done with a Gaussian with FWHM$\approx$100 arcsec and the individual
measurements are also weighted according to their error estimates. The
calculated values are plotted against the 180\,$\mu$m surface brightness and
a straight line is fitted to the points with a least squares algorithm that
takes into account the error estimates on both axes. The procedure is
repeated for all maps and the slopes of the fitted lines provide the
emission spectrum i.e.\ for each observed wavelength the relative intensity
with respect to the 180\,$\mu$m emission.

\begin{figure}[ht]
\begin{center}
\leavevmode
\centerline{\epsfig{file=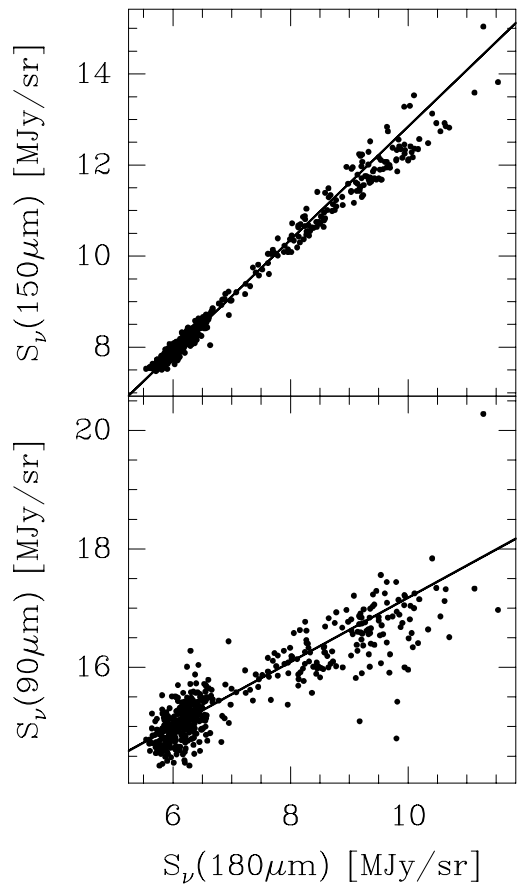, width=6.7cm}}
\end{center}
\caption{\em Plots of the 90\,$\mu$m and the 150\,$\mu$m surface brightness values
in the EBL26 field as function of 180\,$\mu$m surface brightness. The
fitted least squares lines are also shown. The slopes of these lines give
the ratio between surface brightnesses at different wavelengths i.e.\ the
spectrum of the cirrus emission.}
\label{mj_fig:ebl26_fit}
\end{figure}

Figure~\ref{mj_fig:ebl26_fit} shows as an example the relations between
180\,$\mu$m and 90\,$\mu$m and 180\,$\mu$m and 150\,$\mu$m surface
brightness values in the field EBL26. The fitted least square lines are also
shown. Zodiacal light will appear merely as an offset and does not affect
the slopes that determine the cirrus spectrum. Furthermore, zodiacal light
is weak at 180\,$\mu$m and we are looking only for variations correlated
with 180\,$\mu$m emission.

Galactic cirrus is the dominant source of brightness fluctuations and
therefore the spectra obtained characterize the galactic cirrus emission.
However, in low surface brightness regions there may be a significant
contribution from faint extragalactic sources. The DIRBE (\cite{mj_hauser})
and FIRAS (\cite{mj_fixsen98}) experiments indicated a FIR cosmic infrared
background flux of ~1 MJy sr$^{-1}$ between 100 and 240\,$\mu$m. In fields
like EBL22 this would mean that at 180\,$\mu$m about one third of the
surface brightness is due to extragalactic sources. Since the cirrus spectra
are determined from the brightness variations we must consider the
brightness fluctuations caused by these sources in relation to the cirrus
fluctuations.

The cirrus power spectrum is proportional to $B_0/k^3$, where $B_0$ is the
cirrus brightness and $k$ the spatial frequency (\cite{mj_gautier92}). In
other words, cirrus fluctuations decrease rapidly as we move to fainter
cirrus regions and smaller spatial scales.
\cite*{mj_lagache99} have reported a detection of the extragalactic
background fluctuations at 175\,$\mu$m in the Marano\,1 field that has
cirrus surface brightness comparable to our fields. Even after removal of
detected sources they found at scales below 10 arcmin fluctuations in excess
of the expected cirrus contribution. This was interpreted to be caused by
point sources below the detection limit.

The larger dimension of our maps is typically 1.5 degrees and most of the
surface brightness variations can be expected to be due to cirrus. This is
particularly clear in fields NGPS and EBL26 where there is a clear intensity
gradient along the mapped strip consistent with the larger scale cirrus
distribution visible in IRAS maps. On the other hand, the field EBL22 is
remarkably flat and it is conceivable that in the absence of large cirrus
structures the extragalactic point sources could affect the derived
spectrum. The small dynamic range means, however, that in this case the
accuracy of the derived spectrum is not very good. The fields VCN and VCS
were observed close the bright galaxy NGC\,5907 and the VCS field actually
extends over the centre of the galaxy. Pixels that were clearly affected by
the emission from NGC\,5907 were first discarded but it is still possible
that the derived spectra are affected by emission of the galaxy. Because of
the small field size ($\sim$10 arcmin) the VCN spectrum could be affected by
faint extragalactic sources.

\begin{figure}[ht]
\begin{center}
\leavevmode
\centerline{\epsfig{file=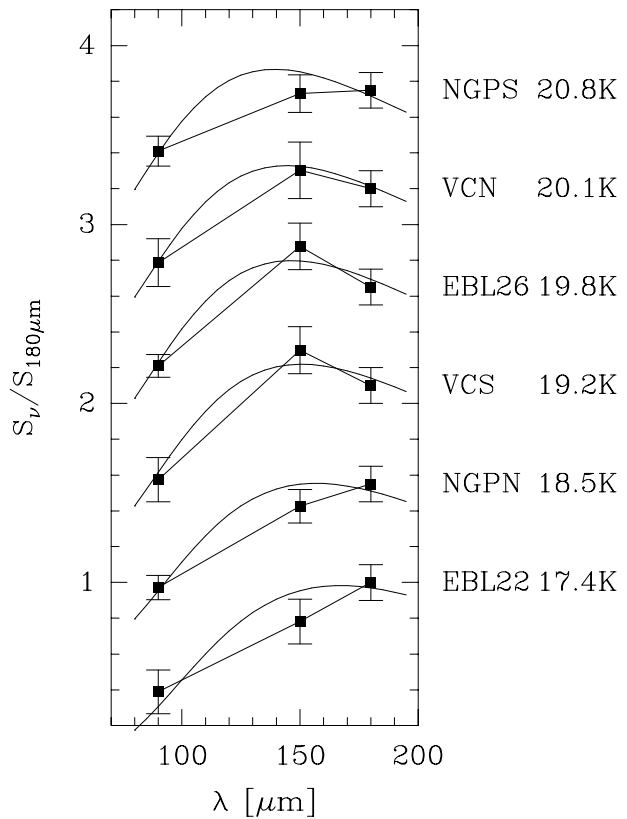, width=8.1cm}}
\end{center}
\caption{\em Cirrus spectra derived from the six observed fields. The
error estimates contain the statistical errors provided by the fitting
procedure and a further 10\% error representing the uncertainty of the
relative calibration between the filters. The solid lines show the fitted
modified black body curves, $\nu^2 B_{\nu}(T)$. }
\label{mj_fig:cirrus}
\end{figure}

The spectra obtained for the six fields are presented in
Figure~\ref{mj_fig:cirrus}. The error bars include the error estimates
obtained from the fitting procedure and a further 10\% error representing
the error in the relative calibration of the observations made with
different filters. Modified Planck curves, $\nu^2 B_{\nu}(T)$, were fitted
to the data and the dust temperatures obtained are shown in the figure.

\section{DISCUSSION}

We have determined cirrus dust temperatures in six fields with low surface
brightness. The derived dust temperatures, $\sim$18-20\,K, are higher than the
typical temperatures estimated for galactic dust from  DIRBE and FIRAS
experiments assuming the same $\nu^2$ emissivity law. However, the values
are still within the range found in these studies.
\cite*{mj_lagache98} found an average temperature of 17.5\,K while higher
temperatures were mostly associated with star forming regions. Temperatures
exceeding 20\,K were seen only towards the galactic centre and some
prominent star forming regions although the scatter in the temperatures is
more than 1\,K at all galactic latitudes.

Our fields are all at high galactic latitudes (see
Table~\ref{mj_table:maps}). Based on COBE measurements at 100--240\,$\mu$m
\cite*{mj_boulanger96} derived average dust temperatures of $T=17.5$\,K for
regions at $|b|>$30 degrees.  Earlier FIRAS estimates (\cite{mj_reach95})
also suggested similar or lower values. However, based on the DIRBE
140\,$\mu$m to 240\,$\mu$m ratios \cite*{mj_dwek97}) reported
colour temperatures of $T\sim$19\,K for regions with $|b|>$45 degrees.

In Section~\ref{mj_sect:cal} we found systematic differences between ISOPHOT
and DIRBE calibrations. Compared with DIRBE the ISOPHOT surface brightnesses
were found to be higher at 90\,$\mu$m and lower at longer wavelengths.
According to Table~\ref{mj_table:dirbe} adopting the DIRBE calibration would
increase our 90\,$\mu$m data by up to 30\% relative to the longer
wavelengths. This would reduce the estimated dust temperature by
$\sim$1.5\,K. Therefore the temperature difference between DIRBE and our
results can be explained entirely by the difference in calibration.

Our temperature determinations are based on data at three wavelengths:
90\,$\mu$m, 150\,$\mu$m and 240\,$\mu$m. The emission at wavelengths
$\lambda>$100\,$\mu$m can be explained by classical grains in equilibrium
with the interstellar radiation field. Smaller grains that are transiently
heated to higher temperatures are responsible for the emission at shorter
wavelengths. Based on both observations and theoretical calculations the
contribution from transietly heated particles is significant already at
60\,$\mu$m and may extend even to slightly longer wavelengths
(\cite{mj_sodroski94}; \cite{mj_dwek97}; \cite{mj_dbp}). The ISOPHOT
90\,$\mu$m filter is wide enough to pick up some radiation from wavelenths
down to $\sim$60\,$\mu$m and our observations may be affected to a small
degree by this other grain population. Our temperature estimates do not
correspond exactly to the equilibrium temperature of large dust particles
but the bias towards higher temperatures is expected to be small.

In the low surface brightness regions the accuracy of DIRBE measurements is
not very good and one can obtain only the average spectrum of a very large
area. Because of the better resolution and sensitivity of ISOPHOT we have
been able to study the cirrus spectrum in separate, small regions. Adopting
the ISOPHOT calibration the accuracy of the temperature estimates is
$\sim$1\,K. For example, the 2\,K difference between the adjacent fields
NGPS and NGPN is likely to be true.

\section*{ACKNOWLEDGMENTS}

The work was supported by the Academy of Finland Grant no. 1011055
and funding from DLR and MPG in Germany.


\end{document}